\documentclass[journal]{IEEEtran}
\usepackage{marvosym}
\usepackage{textcomp}
\usepackage{xcolor}
\usepackage{graphicx}
\usepackage{amsmath}
\usepackage{amssymb}
\usepackage{amsfonts}
\usepackage{mathtools}
\usepackage{epsfig}
\usepackage{epstopdf}
\usepackage{hhline}
\usepackage{subfigure}
\usepackage{stfloats}
\usepackage{cite}
\usepackage{algorithm}
\usepackage{algorithmic}
\usepackage[CJKbookmarks=true,
bookmarksnumbered=true,
bookmarksopen=true]{hyperref}
\usepackage{multirow}
\usepackage{diagbox}

\begin{document}
\title{Lightweight Neural Network with Knowledge Distillation for CSI Feedback}
\author{Yiming~Cui,
Jiajia~Guo,
Zheng~Cao,
Huaze~Tang,
Chao-Kai~Wen, \emph{Fellow}, \emph{IEEE},\\
Shi~Jin, \emph{Fellow}, \emph{IEEE},
Xin Wang, \emph{Member}, \emph{IEEE}, 
and Xiaolin Hou, \emph{Senior Member}, \emph{IEEE}
	
\thanks{

Y. Cui, J. Guo, Z. Cao, H. Tang, and S. Jin are with the National Mobile Communications Research Laboratory, Southeast University, Nanjing 210096, China (e-mail: cuiyiming@seu.edu.cn; jiajiaguo@seu.edu.cn; zheng\_cao@seu.edu.cn; hztang@seu.edu.cn; jinshi@seu.edu.cn).
		
C.-K. Wen is with the Institute of Communications Engineering, National Sun Yat-sen University, Kaohsiung 80424, Taiwan (e-mail: chaokai.wen@mail.nsysu.edu.tw).

X. Wang and X. Hou are with DOCOMO Beijing Communications Laboratories Co., Ltd., Beijing, P. R. China (e-mail: wangx@docomolabs-beijing.com.cn, hou@docomolabs-beijing.com.cn).}

\thanks{An earlier version of this paper was presented at the IEEE
VTC2021-Fall \cite{tang2021knowledge}.}
}

\maketitle

\vspace{-2cm}
\begin{abstract}

Deep learning has shown promise in enhancing channel state information (CSI) feedback. However, many studies indicate that better feedback performance often accompanies higher computational complexity. Pursuing better performance-complexity tradeoffs is crucial to facilitate practical deployment, especially on computation-limited devices, which may have to use lightweight autoencoder with unfavorable performance. To achieve this goal, this paper introduces knowledge distillation (KD) to achieve better tradeoffs, where knowledge from a complicated teacher autoencoder is transferred to a lightweight student autoencoder for performance improvement. Specifically, two methods are proposed for implementation. Firstly, an autoencoder KD-based method is introduced by training a student autoencoder to mimic the reconstructed CSI of a pretrained teacher autoencoder. Secondly, an encoder KD-based method is proposed to reduce training overhead by performing KD only on the student encoder. Additionally, a variant of encoder KD is introduced to protect user equipment and base station vendor intellectual property. Numerical simulations demonstrate that the proposed methods can significantly improve the student autoencoder's performance, while reducing the number of floating point operations and inference time to 3.05\%--5.28\% and 13.80\%--14.76\% of the teacher network, respectively. Furthermore, the variant encoder KD method effectively enhances the student autoencoder's generalization capability across different scenarios, environments, and bandwidths.

\end{abstract}


\section{Introduction}

Massive multiple-input multiple-output (MIMO) is a crucial technology in fifth generation (5G) communication systems and is expected to be further advanced in sixth generation communication systems \cite{boccardi2014five}. The base station (BS) in massive MIMO typically has a large number of antennas to achieve advantages such as higher spectral and energy efficiency. However, in frequency division duplex (FDD) massive MIMO systems, downlink channel acquisition has long been a challenging problem due to the lack of channel reciprocity. This means that downlink channel acquisition is dependent on channel state information (CSI) feedback from user equipment (UE) \cite{marzetta2010noncooperative}. The large number of antennas results in substantial feedback overhead, which becomes a major obstacle to FDD massive MIMO \cite{liang2016fdd}. An efficient CSI feedback method is essential to fully realize the benefits of massive MIMO.

To address the significant burden of feedback overhead, various CSI feedback techniques have emerged during the last few decades. These techniques are primarily categorized into three groups: codebook-based \cite{love2008overview}, compressive sensing (CS)-based \cite{kuo2012compressive}, and deep learning (DL)-based CSI feedback \cite{guo2022overview}. In codebook-based feedback, the UE selects a codeword from a shared codebook with the BS and communicates the index back to the BS. The BS then retrieves the corresponding codeword. This method, including the Type I/II codebook in 5G New Radio (NR) \cite{TYPE12,ghosh20185g}, is standardized but limited to regularly shaped antenna arrays and becomes complex with more antennas. In CS-based feedback, CSI is compressed using random projection and then restored through algorithms like BM3D-AMP \cite{metzler2016denoising,sim2016compressed}. However, these algorithms are computationally demanding and may not be feasible in real-world systems. Additionally, the effectiveness of CS-based feedback heavily relies on the CSI's sparsity, which is not always present in actual scenarios.

Recently, DL has achieved considerable success in many areas, such as computer vision. This motivates researchers to apply DL to communications for performance improvement, especially in the physical layer. Typically, certain modules in the physical layer can be designed with neural networks including channel prediction \cite{booth2020deep,yang2020deep}, channel mapping \cite{alrabeiah2019deep}, beamforming design \cite{soltani2019deep,mashhadi2021pruning}, and more. To overcome the limitations of codebook-based and CS-based methods and improve feedback performance, DL was first introduced into CSI feedback in \cite{wen2018deep}. This method leverages an autoencoder neural network, called CsiNet, to automatically learn CSI compression and reconstruction from a large number of CSI samples. Due to the ability to extract environment knowledge, DL-based CSI feedback significantly outperforms codebook-based and CS-based methods, and is widely recognized as a promising technology in future communication systems. In recent years, intensive research has been conducted on DL-based CSI feedback, focusing on feedback performance improvement (or equivalently feedback overhead reduction), joint design with other modules, and deployment considerations.

To improve feedback performance, novel neural network architectures are introduced to design autoencoder networks for CSI feedback, including attention mechanism \cite{cai2019attention,zhang2022attention,liu2022model}, multiple-resolution blocks \cite{lu2020multi,hu2021mrfnet}, and generative networks \cite{tolba2020massive,hussien2022prvnet}. Additionally, different types of correlations are leveraged to reduce feedback overhead, such as temporal correlation \cite{lu2018mimo,wang2018deep}, bi-directional correlation \cite{liu2021hyperrnn}, adjacent frequency channel correlation \cite{wang2021compressive}. These works extend the basic CsiNet-like framework, further demonstrating the high potential of DL-based CSI feedback. Furthermore, DL-based CSI feedback is also jointly designed with other modules including precoding \cite{wu2022deep}, channel estimation \cite{sun2020ancinet}, and pilot training \cite{liu2021hyperrnn}. In 2021, DL-based CSI feedback has been selected as a representative use case in a study item for NR air interface by the 3rd generation partnership project (3GPP) Release 18 \cite{213599}.\footnote{Anothor two representative use cases are beam management \cite{xue2023beam} and positioning accuracy enhancements \cite{wang2017csi}.} The focus of related research has gradually shifted from pursuing improved feedback performance to addressing deployment challenges. These problems are widely studied, including bit-stream generation \cite{ravula2021deep}, multiple-rate feedback \cite{wang2021novel}, imperfect feedback \cite{ye2020deep}, and more.

In addition to the aforementioned research, computational limitations are also considered to facilitate the deployment of DL-based CSI feedback in practical communication systems. While parallel-computing architectures, such as graphics processing units (GPUs), can significantly accelerate DL-based CSI feedback algorithms, the computational complexity may still be too high for resource-limited devices with strict inference delay requirements \cite{guo2020compression}. To address this issue, previous research has mainly focused on empirical lightweight network architecture designs and other universal network compression methods. For empirical architecture designs, lightweight autoencoder networks such as ConvSquCsiNet \cite{guo2020compression}, CLNet \cite{ji2021clnet}, and STNet \cite{mourya2022spatially} have been proposed. These variants of CsiNet can achieve satisfactory performance with relatively small complexity. In addition, universal network compression methods have been introduced into DL-based CSI feedback \cite{guo2020compression}, including network pruning, low-rank factorization, parameter quantization/binarization.

By optimizing the network architectures and weight structures, the aforementioned methods effectively reduce the complexity of autoencoder networks with moderate performance loss, making it possible to implement high-performance autoencoder networks on computation-limited devices. However, this performance-complexity tradeoff can be further improved even when the network architectures and weight structures are fixed. Notably, more complex autoencoder networks usually perform better and can learn more intricate environmental knowledge, which simpler networks cannot. In recent years, a novel training strategy called knowledge distillation (KD) has been proposed to achieve neural networks with high performance and low complexity \cite{hinton2015distilling,gou2021knowledge,phuong2019towards}. The key idea of KD is that the knowledge learned by one or an ensemble of complicated teacher networks can be transferred into a simple student network, which is realized by forcing the simple student networks to imitate the outputs of the complicated networks. Then, the student networks can achieve higher performance compared to being trained with ordinary training methods.

In this paper, we introduce KD-aided neural network lightweight methods to DL-based CSI feedback. These methods are compatible with different neural network architectures and can be used with other common neural network lightweight methods to further improve the performance-complexity tradeoff. We propose an autoencoder KD method where a pretrained teacher autoencoder is used to distill knowledge to a lightweight student autoencoder, leading to substantially improved performance. Given that training a whole student autoencoder from scratch incurs significant training overhead, we propose an encoder KD method. The teacher encoder network is used to distill knowledge to the lightweight student encoder network, and the teacher decoder network is directly used to avoid training the decoder from scratch. Notably, all the aforementioned methods involve sharing encoder architectures between the BS and UE, which can cause intellectual property problems between different vendors. To avoid this problem, we propose a variant of the encoder KD method that avoids transmitting encoder architectures between the BS and UE. Finally, we validate the performance of the proposed lightweight strategies with CSI samples generated by QuaDRiGa \cite{jaeckel2014quadriga}.

The contributions of this paper can be summarized as follows:
\begin{itemize}

\item \textbf{Improved performance}: We introduce KD to DL-based CSI feedback for neural network lightweight. An autoencoder KD-based method is proposed by letting the student network imitate the outputs of the teacher network. The feedback performance of the lightweight autoencoder can be significantly improved after autoencoder KD.

\item \textbf{Reduced training overhead}: To reduce the training overhead in the autoencoder KD-based method, we propose an encoder KD-based method. The feedback performance is also improved compared to the autoencoder KD-based method.

\item \textbf{Protecting intellectual property}: We propose a variant of the encoder KD-based method, where the encoder and the decoder can be trained separately without knowledge of the entire autoencoder architecture, thereby protecting the intellectual property of different vendors.

\item \textbf{Improved generalization capability}: The variant encoder KD-based method can also be leveraged to improve the generalization capability of the student autoencoder in different environments, scenarios, or settings.
\end{itemize}

The organization of the following sections is shown as follows. The system model is introduced in Section II. The autoencoder KD-based lightweight method is introduced in Section III. The encoder KD methods are subsequently proposed in Section IV. The simulation results of the proposed methods with QuaDRiGa channel datasets are presented in Section V. The conclusion of this research is drawn in Section VI.

The notations in this paper are listed as follows. The unbold lowercase letters, boldface lowercase letters, and boldface uppercase letters represent scalars, vectors, and matrices, respectively. $\mathbb{C}^{\rm s\times \rm t}$ denotes an complex space sized $\rm s$ by $\rm t$. $(\cdot)^H$ denotes Hermitian transpose. $\|\cdot\|_2$ donotes Euclidean norm.

\section{System Model}
The system model is introduced in this section. First, the channel model and signal model are presented. Then, the framework of DL-based CSI feedback is introduced.

\subsection{Channel Model and Signal Model}
A single-cell FDD massive MIMO system is considered. A uniform linear array (ULA) is adopted at the BS, where the number of antennas is $N_{\rm t} \gg 1$. The UE is equipped with a single antenna. Orthogonal frequency division multiplexing (OFDM) is employed, where the number of subcarriers is $N_{\rm c}$. For the $n$-th subcarrier, the channel can be represented as
\begin{equation}
    \tilde{\mathbf{h}}_n=\sum_{c=1}^{N_{\rm c}}\sum_{s=1}^{N_{\rm s}}g_{c,s,n}\mathbf{a}({\theta_{c,s}}),
\end{equation}
where $N_{\rm c}$, $N_{\rm s}$, $g_{c,s,n}$, and $\theta_{c,s}$ represent the number of clusters, the number of subpaths for each cluster, the complex gain of the $s$-th subpath in the $c$-th cluster, and the angle-of-departure of this subpath, respectively. The steering vector $\mathbf{a}({\theta})$ can be formulated as
\begin{equation}
    \mathbf{a}({\theta})=[1,e^{j 2 \pi \frac{d}{\rm \lambda}{\rm sin}(\theta)},...,e^{j 2 \pi \frac{(N_{\rm t}-1)d}{\rm \lambda}{\rm sin}(\theta)}],
\end{equation}
where $d$ is the distance between adjacent antennas and $\lambda$ denotes the wavelength of each carrier. Then, the received signal $y_n \in \mathbb{C}$ for the $n$-th subcarrier can be formulated as
\begin{equation}
    y_n={\Tilde{\mathbf{h}}}_n^H \mathbf{v}_n x_n + z_n,
\end{equation}
where $\mathbf{v}_n \in \mathbb{C}^{N_{\rm t}\times 1}$, $x_n \in \mathbb{C}$, and $z_n \in \mathbb{C}$ denotes the beamforming vector, the information-bearing symbol, and additive Gaussian white noise, respectively. The channel vectors ${\Tilde{\mathbf{h}}}_n$ can be stacked together as the CSI matrix in spatial-frequency domain $\tilde{\mathbf{H}}=[{\Tilde{\mathbf{h}}}_1,...,{\Tilde{\mathbf{h}}}_{N_{\rm c}}] \in \mathbb{C}^{N_{\rm t}\times {N}_{\rm c}}$. The CSI matrices are usually sparse in angular-delay domain. Therefore, discrete Fourier transform (DFT) is performed to the CSI matrix $\tilde{\mathbf{H}}$ to achieve the CSI in angular-delay domain, which can be formulated as follows:
\begin{equation}
    \mathbf{H}= \mathbf{F}_{\rm a}\tilde{\mathbf{H}}\mathbf{F}_{\rm d},
\end{equation}
where $\mathbf{F}_{\rm a} \in \mathbb{C}^{N_{\rm t} \times N_{\rm t}}$ and $\mathbf{F}_{\rm d} \in \mathbb{C}^{N_{\rm c} \times N_{\rm c}}$ are DFT matrices.\footnote{In practice, DL-based CSI feedback is not limited to ULA and can be used with arbitrary arrays. In this work, ULA is adopted for convenience.}

\subsection{DL-based CSI feedback}

\begin{figure}[t]
    \centering
    \includegraphics[width=0.45\textwidth]{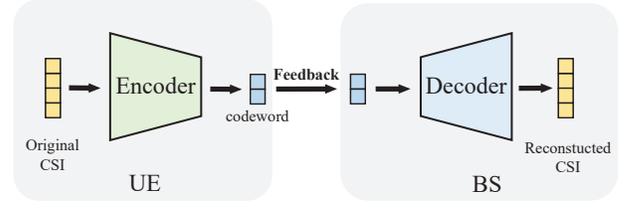}
    \caption{Illustration of DL-based CSI feedback.}
    \label{fig:csifeedback}
\end{figure}

Transmitting the CSI matrix $\mathbf{H}$ to the BS causes overwhelming communication overhead. To resolve this problem, DL-based CSI feedback is introduced to CSI feedback to alleviate feedback overhead. As shown in Fig. \ref{fig:csifeedback}, an autoencoder network is adopted, which is composed of an encoder and a decoder. When UE completes channel estimation and acquires the downlink CSI, the CSI is compressed by the encoder network into a real-valued codeword vector $\mathbf{s}$ as follows:
\begin{equation}
    {\rm \mathbf{s}}=f_{\rm en}(\mathbf{H}),
\end{equation}
where $f_{\rm en}(\cdot)$ denotes the encoder network and the length of the codeword $\mathbf{s}$ is $N_{\rm s}$. Channel estimation error is not considered in this paper. The compression ratio $\gamma$ is formulated as follows:
\begin{equation}
    \gamma=\frac{N_{\rm s}}{2N_{\rm t}N_{\rm c}}.
\end{equation}
Then, the UE transmits the codeword to the BS through a feedback link. Perfect transmission is assumed for the feedback link. When the BS receives the codeword, the codeword is fed into the decoder to reconstruct the CSI as follows:
\begin{equation}
    \mathbf{\hat{\mathbf{H}}}=f_{\rm de}({\rm \mathbf{s}}),
\end{equation}
where $\hat{\mathbf{H}}$ denotes the reconstructed CSI and  $f_{\rm de}(\cdot)$ represents the decoder network. Finally, the CSI $\hat{\mathbf{H}}$ is transformed back into spatial-frequency domain by inverse DFT.

The primary goal of DL-based CSI feedback is the accurate recovery of each element in the entire channel matrix. Mean square error (MSE), being an intuitive loss function,
\begin{equation}
    l_{\rm{MSE}}(\mathbf{H},\hat{\mathbf{H}})=\|\hat{\mathbf{H}}-\mathbf{H}\|_{2}^{2},
\end{equation}
is widely used in related works to train the autoencoder network.

\section{Autoencoder KD for CSI Feedback}
In this section, we will begin by introducing the motivation behind the autoencoder KD-based lightweight method. Following this, we will discuss the basic autoencoder architectures and provide elaboration on the details of the autoencoder KD method.

\subsection{Motivation}
In recent years, a number of variants of CsiNet have been proposed to improve feedback performance. As summarized in \cite{guo2022overview}, it is common for the improvement of feedback performance to come with an increase in complexity. To address this issue, several methods have been proposed to implement a lightweight neural network, which can be broadly categorized into three categories: optimization of neural network architectures, optimization of neural network weights, and optimization of training strategies. For DL-based CSI feedback, the optimization of neural network architectures and weights has been introduced before, such as efficient structural design, pruning, quantization, and binarization \cite{guo2020compression}. However, optimization of training methods for lightweight has not been introduced to DL-based CSI feedback. KD, as a representative training strategy optimization method, can be combined with the aforementioned lightweight methods to further improve their performance. Therefore, in this section, we introduce a KD-based autoencoder lightweight method for DL-based CSI feedback.

\begin{figure*}[t]
    \centering
    \includegraphics[width=0.8\textwidth]{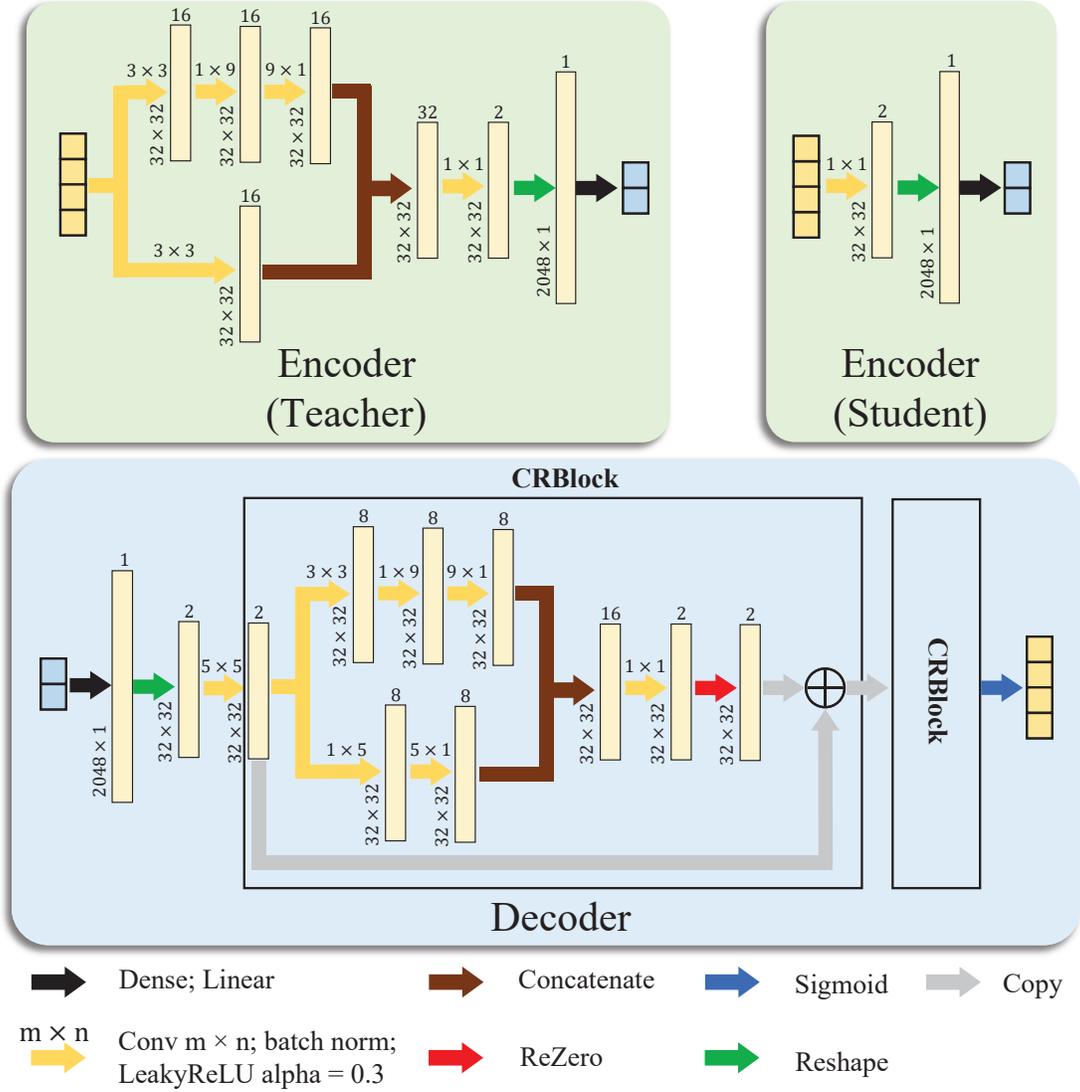}
    \caption{Illustration of the autoencoder architectures adopted in this research.}
    \label{fig:csiarch}
\end{figure*}

\subsection{Key Idea and Training Framework}

The deployment of autoencoder KD-based neural network lightweight can be summarized into the following three steps: First, the BS pretrains a complicated autoencoder as the teacher network. Second, the BS trains the lightweight student network with the supervision of both the ground-truth CSI and the CSI reconstructed by the teacher autoencoder. Finally, the lightweight encoder of the student autoencoder is transmitted to the UE for deployment. We will provide implementation details and in-depth explanations for these steps in the following subsections.

1) Basic Autoencoder Architectures for KD. As mentioned above, a teacher autoencoder network and a student autoencoder network are required to implement KD. The basic architectures of the adopted teacher network and student network are first introduced. In DL-based CSI feedback, the BS is typically has adequate computational resources, while UE such as mobile phones and other IoT devices have limited resources. Therefore, the decoder network of the student autoencoder can have a complicated architecture to achieve the highest possible performance, while the encoder network should have a lightweight architecture. To implement KD, we use a complicated teacher autoencoder network called CRNet, and a student autoencoder network called CRNet-SE with a simplified encoder. The architectures of the two autoencoder networks are modified based on \cite{lu2020multi} and \cite{wen2018deep}, and are illustrated in Fig. \ref{fig:csiarch}. Both CRNet and CRNet-SE adopt the same decoder, but their encoder networks are different. The encoder of the teacher network CRNet is complicated to extract knowledge prepared for KD, while the encoder of the student network CRNet-SE is designed to be lightweight for deployment in UEs. The architectures of these two autoencoder networks are elaborated as follows.

The CSI matrices are divided into real and imaginary parts as two channels for input. For CRNet-SE, the encoder network is composed of a convolutional layer for feature extraction and a dense layer for compression. Compared with CRNet-SE, the encoder of CRNet adopts a more complicated feature extraction module with a multiple-resolution structure. The input CSI is processed with two branches of convolutional layers with different kernel sizes. The output of the two branches are then concatenated and merged into the original size of CSI for compression.

The decoder is mainly composed of a dense layer, a head convolutional layer, and two CRBlocks. The CRBlock adopts a similar multiple-resolution structure with the encoder of CRNet. In each CRBlock, skip connection \cite{he2016deep} and a ReZero \cite{bachlechner2021rezero} module are added in each CRBlock to improve the convergence. Finally, a sigmoid function is added at the end of the decoder to ensure that the output values are between 0 to 1. All convolutional layers are followed by a batch normalization layer and a LeakyReLU activation ($\alpha=0.3$), except for the last convolutional layers in CRBlock.

The details of the teacher and student autoencoder are annotated in Fig. \ref{fig:csiarch}. The arrows in different colors represent different types of layers and operations and the yellow rectangles denote the output of each layer. The annotations on the arrows represent the sizes of the convolutional kernels. The annotations on the left of the layer output are the length and width of the output and the ones on the top mean the channel numbers.

\begin{figure}[t]
	\centering
	\subfigure[\label{fig_first_case1}Pretraining a teacher network]{\includegraphics[width=1\linewidth]{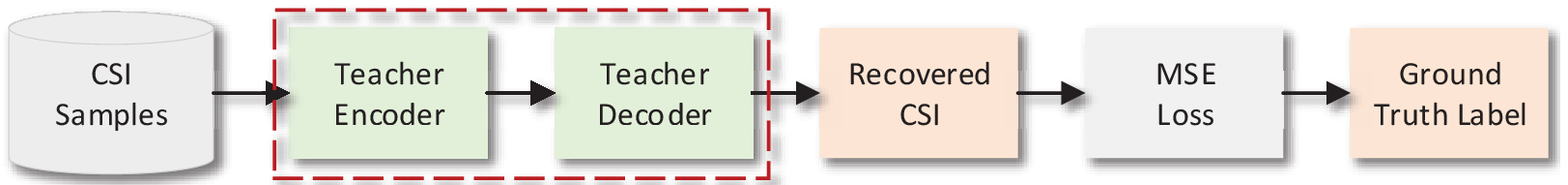}}
	\subfigure[\label{fig_second_case1}Autoencoder distillation]{\includegraphics[width=1\linewidth]{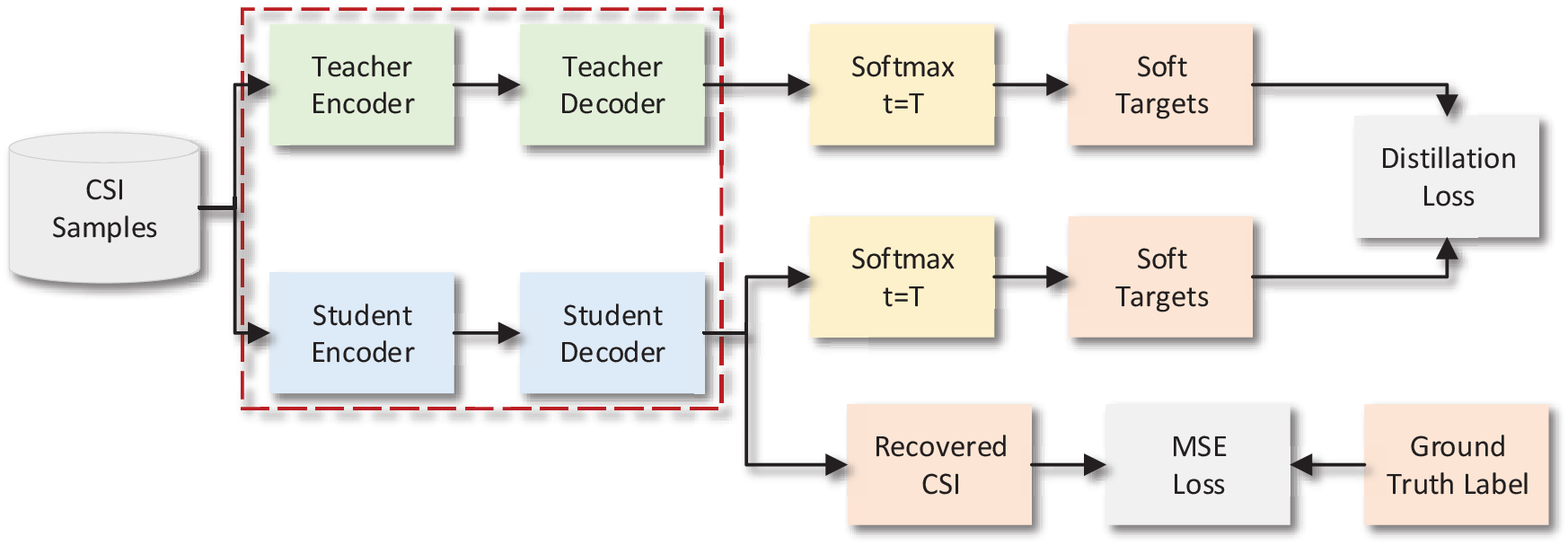}}
	\caption{Illustration of autoencoder KD-based neural network lightweight for CSI feedback.}
	\label{fig:kd}
	\vspace{-0.6cm}
\end{figure}

2) Knowledge and Dark Knowledge. In DL-based CSI feedback, a large number of CSI samples are fed into an autoencoder network to extract environment knowledge. This knowledge is usually viewed as parameter values of autoencoder networks, which can make inter-model knowledge transferring seem difficult. However, from the perspective of \cite{hinton2015distilling}, the knowledge can also be regarded as the mapping from network inputs to outputs. This perspective facilitates the idea that the form of knowledge can be changed without modifying the knowledge itself. The knowledge learned by a complicated network can be transferred to a simple network by letting the simple network imitate the outputs of the complicated network. In this case, the form of knowledge is converted to a simpler one while the knowledge itself is kept the same. This approach can help avoid using cumbersome networks in actual deployment.

Compared to direct learning with labels\footnote{The DL-based CSI feedback algorithm is also a type of self-supervised learning, hence the ground-truth CSI used to calculate the loss function at the autoencoder outputs is referred to as the label in this paper.}, the outputs of the teacher network contain more inconspicuous knowledge, which may be learned by complex networks but is not easily captured by simpler student networks. This type of knowledge is commonly known as dark knowledge. Here, we delve deeper into the explanation of dark knowledge. 
In DL-based CSI feedback, which involves lossy compression, perfect CSI reconstruction at the BS is rarely achievable. Learning with the aid of the teacher autoencoder's output, a feasible sub-optimal solution, is intuitively more beneficial for the optimization process compared to learning directly from the ground-truth CSI, which is essentially an infeasible solution. In other words, the dark knowledge in the teacher autoencoder's output additionally indicates the degree of accuracy achievable in CSI reconstruction at a certain compression ratio. To enhance the efficiency of learning dark knowledge, an extended softmax function is introduced \cite{hinton2015distilling}, formulated as follows:
\begin{equation}
    \sigma(\mathbf{z};t)=\frac{{\rm exp}(\mathbf{z}/t)}{\sum_{i}{\rm exp}(z_{i}/t)},
\label{equ:softmaxt}
\end{equation}
where $\mathbf{z}$, $z_i$, and $t$ represent the outputs of the teacher network, $i$-th element in the outputs of the teacher network, and a hyper-parameter called temperature, respectively. The extended softmax function is reduced to the ordinary softmax function when $t=1$. The outputs of the extended softmax function are also called soft targets.

As the temperature $t$ increases, the imperceptible small values in the CSI, which may contain dark knowledge, are further enlarged, and the large values are weakened. An appropriate value of $t$ makes the dark knowledge in the outputs of the teacher network more evident without destructing other knowledge, and the student network can better learn different knowledge. However, when the $t$ is overlarge, the outputs of the extended softmax are almost uniform, resulting in information loss and performance degradation. Therefore, selecting an appropriate value of $t$ is significant, which is further discussed in the simulation part.

3) Loss Function for KD. The proposed KD-based neural network lightweight method is depicted in Fig. \ref{fig:kd}. After the teacher network is trained, the student network is trained using a combination of distillation loss and ordinary MSE loss. For the distillation loss, a cross-entropy loss function is employed to minimize the discrepancy between the soft targets of the teacher autoencoder output and the student autoencoder output. A high temperature parameter is used to enhance the distillation process. The loss function also incorporates conventional MSE loss alongside the distillation loss to counteract any bias present in the teacher network. The KD loss function is formulated as follows: 
\begin{equation}
    L_{\rm KD}=\alpha\mathcal{H}_1(\mathbf{y},\mathbf{z}_s)+(1-\alpha)\mathcal{H}_2(\sigma(\mathbf{z}_t;{t}),\sigma(\mathbf{z}_s;{t})),
\label{equ:kd}
\end{equation}
where $\mathcal{H}_1$ is the MSE loss function between the outputs of the student network and ground truth CSI, $\mathcal{H}_2$ is the distillation loss function between the soft targets of the student network and the soft targets of the teacher network, $\alpha$ is the hyperparameter controlling the balance of two loss functions, $\mathbf{y}$ is the ground truth CSI, and $t$ is the temperature used in the extended softmax function. After the distillation, the parameters of the student encoder are transmitted to the UE for deployment. According to recent discussions in 3GPP, three typical training types have been identified as potential methods for DL-based CSI feedback \cite{38843}. The autoencoder-based training method is compatible with training Type 1, as specified in TR 38.843.

\section{Encoder KD for CSI Feedback}
In this section, we will begin by introducing the motivation behind encoder KD-based lightweight methods. Then, we will discuss the frameworks of two encoder KD methods, including an ordinary encoder KD method and an intellectual property-protecting variant of the ordinary encoder KD method.

\subsection{Motivation}
The KD-based lightweight method proposed in Section III improves the performance of student autoencoder without increasing any computation complexity in the inference stage. However, the complexity of the training stage is still large. The second step includes training a student autoencoder from scratch, resulting in much training time. A more computationally efficient KD method is required for DL-based CSI feedback.

In actual deployment, the BS is usually equipped with adequate resources to handle the inference of complicated decoder models. Therefore, the student autoencoder adopts the same complicated decoder as the teacher autoencoder in the autoencoder KD method. Rather than training a decoder from scratch for the student autoencoder, a more efficient way is to directly copy the decoder parameters of the teacher autoencoder to the student autoencoder. This approach dramatically reduces the training time because only a lightweight encoder needs to be trained.

\subsection{Key Idea and Training Framework}
A computationally efficient KD-based lightweight method named encoder KD is proposed to reduce the training overhead in KD. Compared with autoencoder KD, which involves distilling knowledge to the whole autoencoder, encoder KD only involves distilling the knowledge of the teacher encoder to the student encoder. To distill knowledge to the student encoder, the MSE between the codewords is adopted as follows:
\begin{equation}
    L_{{\rm KD},{\rm en}}=\mathcal{H}_1(\mathbf{s}_{\rm t},\mathbf{s}_{\rm s}),
\label{equ:ekd}
\end{equation}
where $\mathbf{s}_{\rm t}$ and $\mathbf{s}_{\rm s}$ are the output codewords of the teacher encoder and the student encoder, respectively. In (\ref{equ:ekd}), MSE is adopted to minimize the distance between the codeword labels from the teacher encoder and the output of the student encoder. The reason the cross-entropy loss function is not used is that the encoder output, namely the codeword, lacks a ground truth label. Therefore, the cross-entropy loss function alone cannot adequately guide the training of the encoder.

\begin{figure}[t]
	\centering
	\subfigure[\label{fig_first_case1}Pretraining a teacher network]{\includegraphics[width=1\linewidth]{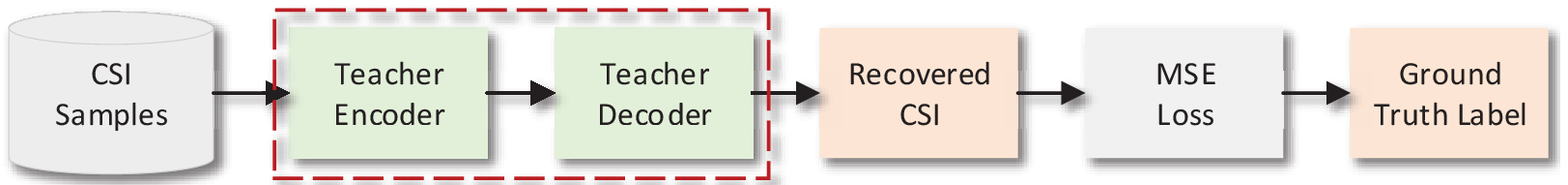}}
	\subfigure[\label{fig_second_case1}Encoder distillation]{\includegraphics[width=0.6\linewidth]{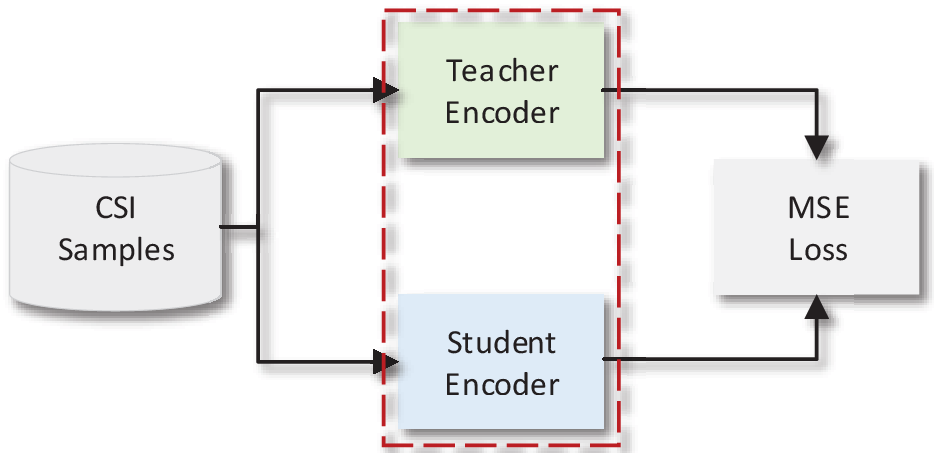}}
	\subfigure[\label{fig_third_case1}Combining and fine-tuning]{\includegraphics[width=1\linewidth]{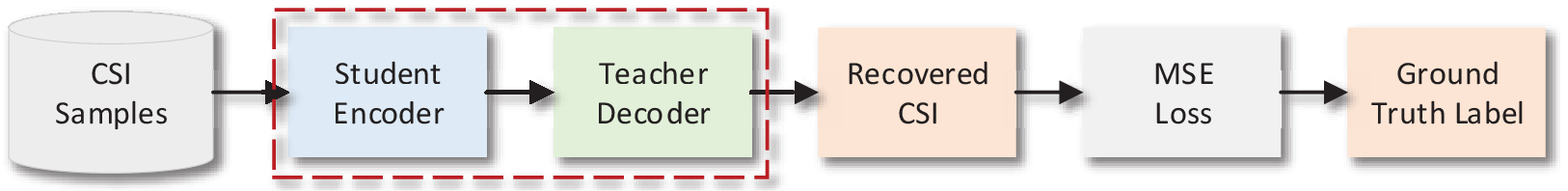}}		
	\caption{Illustration of encoder-distillation for CSI feedback.}
	\label{fig:encoder}
	\vspace{-0.6cm}
\end{figure}

As illustrated in Fig. \ref{fig:encoder}, the encoder KD method comprises three steps, all implemented at the BS as follows:
\begin{itemize}
    \item \textbf{Step 1: Pretraining a Teacher Network---}A well-designed, complex autoencoder is trained to serve as the teacher network. The encoder network is used to distill knowledge into the student encoder. The decoder network is directly employed in the subsequent steps.

    \item \textbf{Step 2: Encoder Distillation---}The BS utilizes the pretrained teacher encoder to distill knowledge into the student encoder using (\ref{equ:ekd}).

    \item \textbf{Step 3: Combining and Fine-Tuning---}The student encoder is concatenated with the teacher decoder to form a new autoencoder network. This whole new autoencoder network is fine-tuned to resolve any mismatches between the student encoder and the teacher decoder. After fine-tuning, the encoder parameters are transmitted to the UE for deployment.
\end{itemize}
Featuring model transfer, the encoder-based training method is compatible with training Type 1, as specified in TR 38.843 \cite{38843}.

Compared with the autoencoder KD-based method, the encoder KD has the following advantages except for the reduction in training time: First, the decoder copied from the teacher autoencoder is usually better than the decoder trained from scratch in the autoencoder KD-based method because the performance of the teacher network is usually the upper bound of the student network after KD. Second, the student encoder can be better trained in encoder KD due to a much simpler back propagation, reducing the risk of gradient vanishing. Therefore, the encoder KD-based lightweight method can bring better performance compared with the autoencoder KD-based method.

\subsection{Intellectual Property-protecting Variant of Encoder KD}

\begin{figure}[t]
	\centering
	\subfigure[\label{fig_first_case1}Pretraining teacher network]{\includegraphics[width=0.95\linewidth]{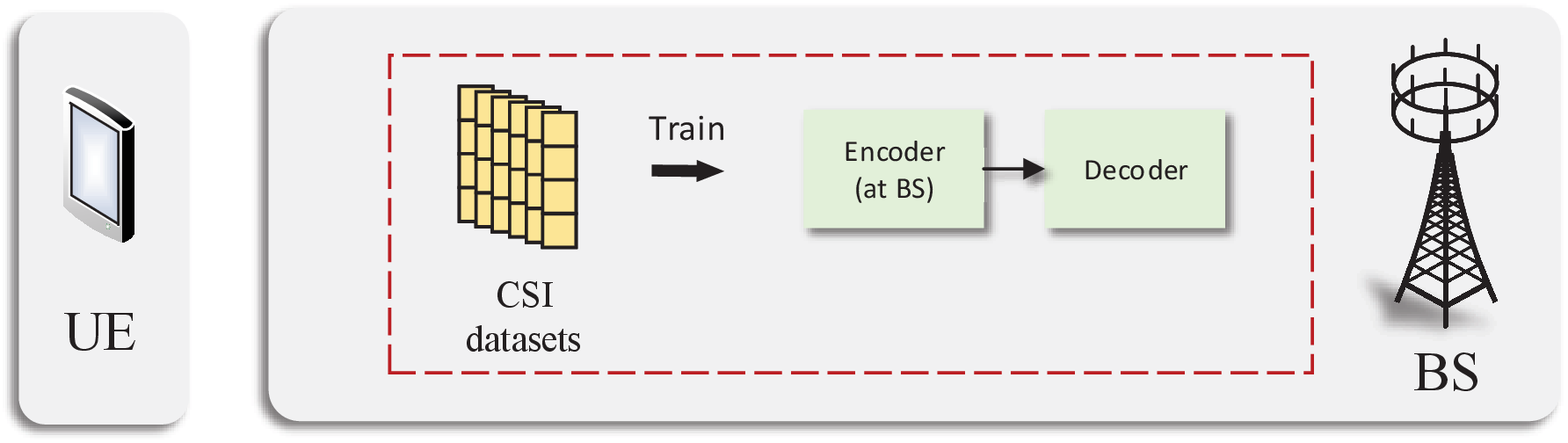}}
	\subfigure[\label{fig_second_case1}Encoder distillation]{\includegraphics[width=0.95\linewidth]{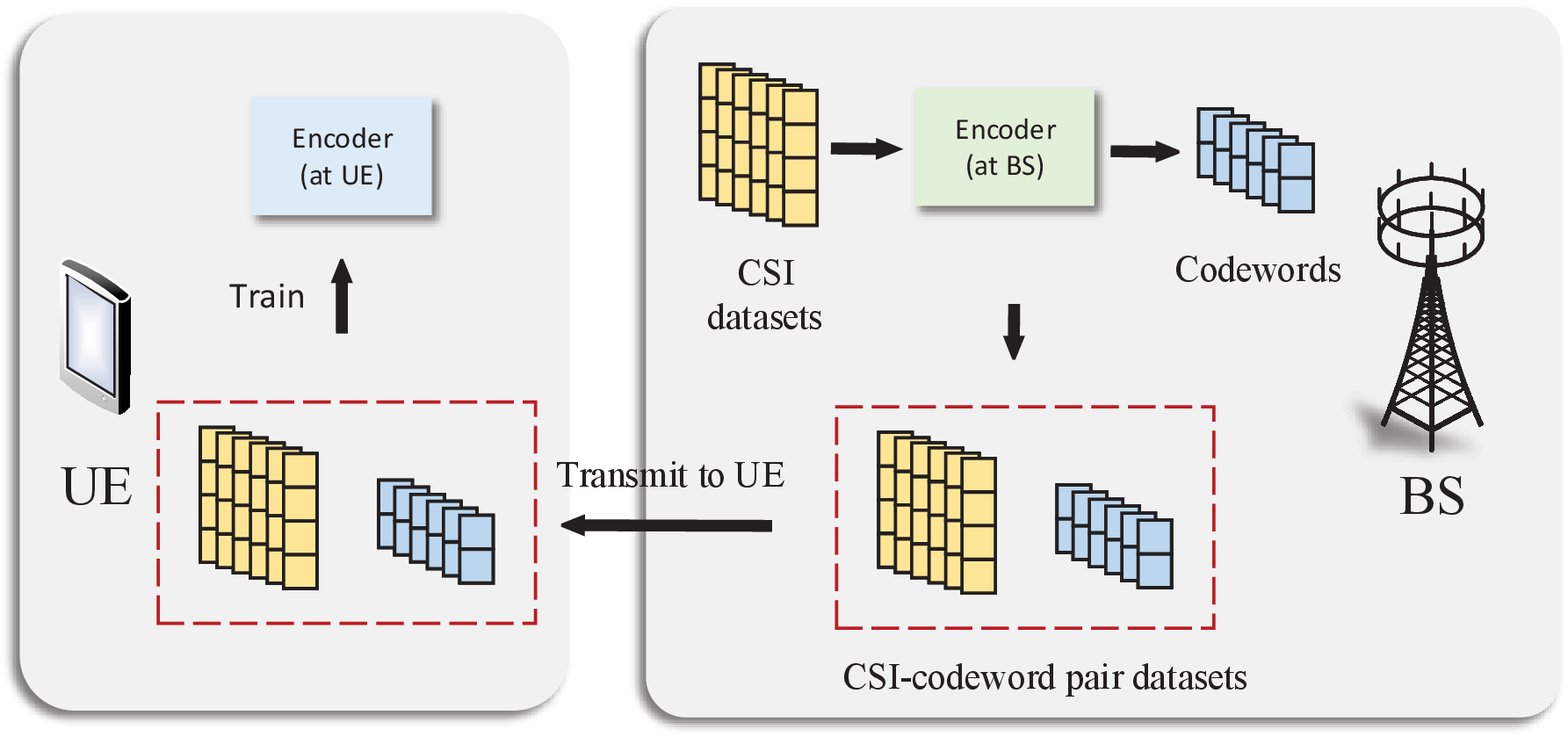}}
	\subfigure[\label{fig_third_case1}Fine-tuning the decoder]{\includegraphics[width=0.95\linewidth]{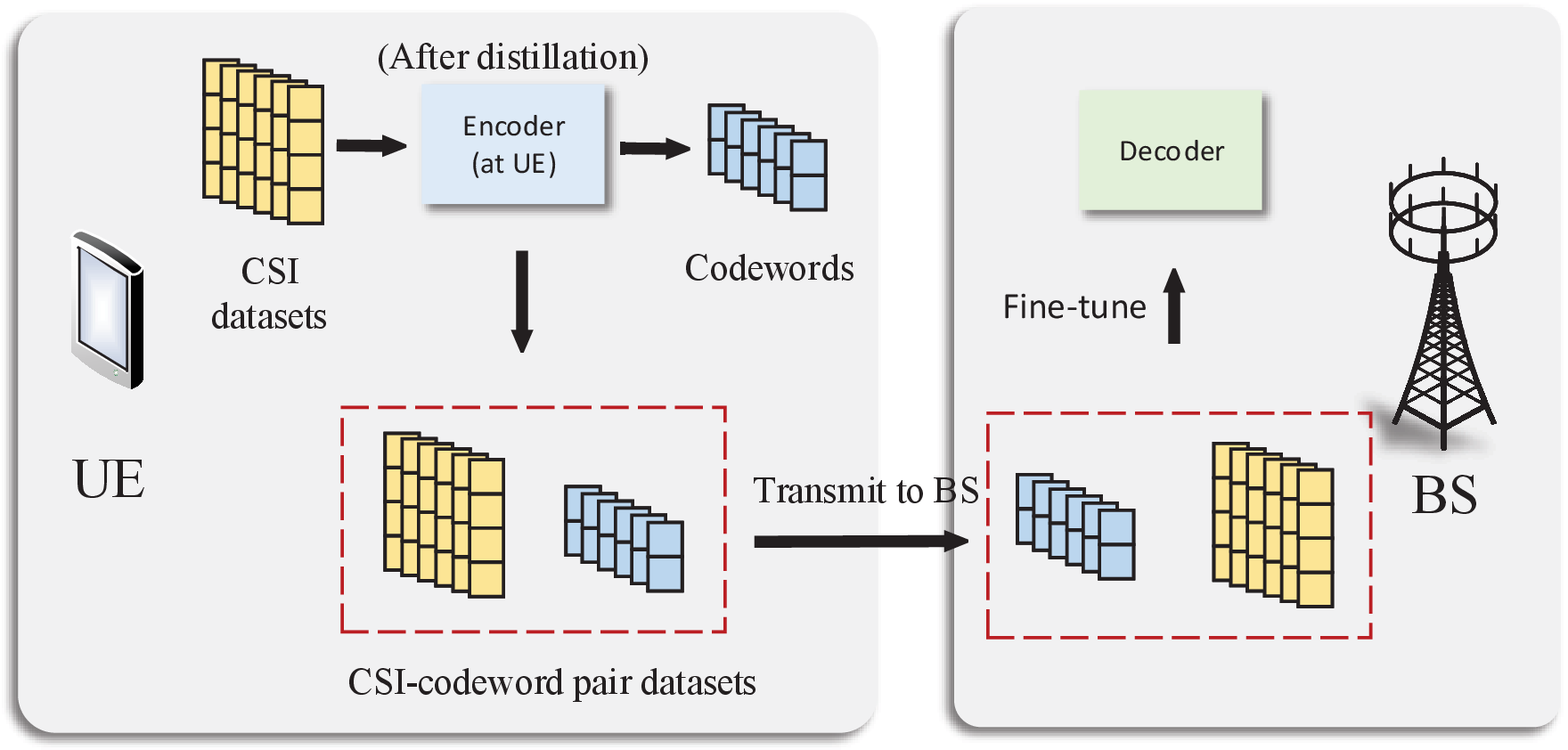}}		
	\caption{Illustration of the intellectual property-protecting encoder KD for CSI feedback.}
	\label{fig:IPPKD}
	\vspace{-0.6cm}
\end{figure}

In the 3GPP discussion regarding the deployment of DL-based CSI feedback, a consensus has emerged among most companies that autoencoder architectures \emph{should not} be standardized, thus allowing each vendor to develop its own neural network. This approach arises from concerns about intellectual property leakage, making model transmission less preferred in practical deployment scenarios. To circumvent this issue, we propose a modified version of the encoder KD method, termed the ``intellectual property-protecting encoder KD.'' This method adapts to situations where the architectures of the encoder and decoder are unknown to each other, which is common in practice.

Typically, to train an autoencoder network, the encoder and decoder are trained together in the same loop. However, when the architectures of these components are kept confidential, this approach becomes infeasible. An alternative strategy is to train the encoder and decoder networks separately using designated input-label pairs, essentially consisting of CSI and codeword data. This method enables independent training of the encoder at the UE and the decoder at the BS, with the training facilitated through the exchange of CSI-codeword pairs. In this scenario, the BS vendor develops a complex teacher encoder and decoder, while the UE vendor creates a lightweight student encoder. The steps of this modified method are visually represented in Fig. \ref{fig:IPPKD} and can be summarized as follows: 
\begin{itemize}
    \item \textbf{Step 1: Pretraining a Teacher Network---}The BS pretrains a teacher autoencoder for encoder KD.

    \item \textbf{Step 2: Encoder Distillation---}The teacher encoder is used to generate a dataset of CSI-codeword pairs. This dataset is then transmitted to the UE, where the UE uses it to train the student encoder with (\ref{equ:ekd}).

    \item \textbf{Step 3: Fine-Tuning the Decoder---}After distillation, the UE uses the student encoder to generate a dataset of CSI-codeword pairs and transmits it to the BS. The BS then uses this dataset to fine-tune the decoder.
\end{itemize}
This training procedure is algorithmically similar to the encoder KD method, with the primary difference being in the fine-tuning phase.

In this method, the BS and UE only share the CSI-codeword datasets and have no knowledge of the network architectures of each other. Therefore, the intellectual properties of the BS vendor and UE vendor are effectively protected. Compared with the encoder KD method proposed in Subsection B, where the whole autoencoder is fine-tuned, in this variant method, the encoder is not fine-tuned in Step 3, which might result in some performance loss. However, considering that the decoder is usually much more complex than the student encoder, this performance loss is negligible. Therefore, the modified Encoder KD method can still provide significant performance gains.

Although sharing large quantities of CSI-codeword pairs may lead to extra communication overhead, this can be remarkably reduced with the following strategies:
\begin{enumerate}

\item \textbf{Designed Parameter Initialization}: As proposed in \cite{chelsea2017model}, initializing the network parameters with a meta-model can reduce the requirement on training samples.

\item \textbf{Data Augmentation}: Data augmentation methods \cite{shorten2019survey} can be employed to enrich the CSI-codeword datasets and enhance feedback performance, thereby reducing the need for CSI-codeword pairs.

\item \textbf{Training Data Quantization}: Quantizing the CSI-codeword pairs using appropriate scalar or vector quantization methods can significantly reduce overhead with minimal performance loss \cite{38843}.
\end{enumerate}

It is important to note that the assumption of the BS vendor's encoder being more complex than the UE vendor's encoder is made solely for the purpose of performing encoder KD. In general, when the two encoder networks are not specified, the modified encoder KD method can be regarded as a separate training method. This allows the UE and BS to train the encoder and decoder independently without sharing network architectures, corresponding to training Type 3 in TR 38.843 \cite{38843}.

In \cite{qualcomm}, a separate training method called sequential training is proposed for DL-based CSI feedback, which includes only Step 2 of the proposed encoder KD method. However, this method does not specify the selection of the encoder architecture. Compared to the sequential training method, our proposed method deliberately introduces a complex teacher network to assist in training the deployed encoder, resulting in higher feedback performance.

\section{Simulation Results}
In this section, we first introduce the channel generation settings and training strategies. We then evaluate the performance of the autoencoder KD method, followed by the validation of the encoder KD method.

\subsection{Simulation Settings}
1) Channel generation settings: QuaDRiGa channel generation software is used to generate the dataset \cite{jaeckel2014quadriga}. The UMi LOS scenario, corresponding to 3GPP TR 38.901 \cite{38901}, is used with a center frequency of 2.655 GHz. The MIMO system bandwidth is set to 70 MHz with 32 subcarriers, and the BS had 32 antennas. Each CSI sample is generated at a random position in the cell. The training set, validation set, and test set contained 100,000, 30,000, and 20,000 samples, respectively. The detailed settings are summarized in Table \ref{table:settings}.

\begin{table}[t]
    \renewcommand{\arraystretch}{1.0}
    \centering
    \caption{Channel Generation Settings}
    \begin{tabular}{l|l}
    \hline
    \hline
         Scenarios & UMi LOS\\
         \hline
         \multirow{2}{*}{Antenna settings}&BS: 32 omnidirectional antennas, ULA \\
         & UE: a single omnidirectional antenna \\
         \hline
         Operating system & FDD-OFDM system \\
         \hline
         Subcarrier number & 32 \\
         \hline
         Center frequency & UMi: 2.655 GHz \\
         \hline
         Bandwidth & 70 MHz\\
         \hline
         Cell range & 100 m \\
         \hline
         BS height & 10 m \\
         \hline
         UE height & 1.5 m\\
         \hline
         Minimum BS-UE distance & 10 m \\
         \hline
         Correlation distance & 12 m \\
    \hline
    \hline
    \end{tabular}
    \label{table:settings}
\end{table}

2)Training settings: TensorFlow 2.4.1 is used to implement the DL-based algorithms on a Intel Xeon E5-2698 CPU and a single NVIDIA Tesla V100 GPU (32 GB graphic memory). An adaptive moment estimation optimizer is used with an initial learning rate of $10^{-3}$, which is reduced to $10^{-4}$ after 100 epochs. The training process is stopped when the best MSE on the validation set no longer decreases for 50 consecutive epochs. The batch size is set to 200. The hyperparameters of KD are set to $T=5$ and $\alpha=0.3$. Normalized MSE (NMSE) is adopted to evaluate the feedback performance as follows:
\begin{equation}
	\rm NMSE= \rm E\left\{\frac{\|\mathbf{H}-\hat{\mathbf{H}}\|_2^2}{\|\mathbf{H}\|_2^2}\right\},
\end{equation}
where ${\rm E}(\cdot)$ denotes expectation. For simplicity, the intellectual property-protecting variant of the encoder KD-based method is referred to as ``Variant encoder KD'' in the results. In addition, training a network from scratch using the MSE loss function without KD is denoted as 'Vanilla' for convenience.

\subsection{Autoencoder KD for CSI Feedback}
\subsubsection{Performance Analysis}
Table \ref{table:kd} summarizes the performance of the teacher network CRNet, a baseline network CRNet-SE with the vanilla training method, and a student network CRNet-SE trained with autoencoder KD. As shown in Table \ref{table:kd}, CRNet-SE trained with autoencoder KD significantly outperforms CRNet-SE with the vanilla training method for different compression ratios. The results demonstrate that CRNet-SE can successfully learn the dark knowledge of CRNet through autoencoder KD. However, the performance of CRNet-SE after autoencoder KD fails to outperform CRNet, implying that the performance of the student autoencoder cannot exceed that of the teacher autoencoder.

\begin{table}[t]
    \vspace{0.7cm}
    \setlength{\abovecaptionskip}{-0.02cm}
    \setlength{\belowcaptionskip}{-0.4cm}	
    \centering
    \caption{NMSE comparison among autoencoder KD with other training strategies}
    {\begin{tabular}[l]{c | c  c  c}					
            \hline
            \hline
            CR & Networks & Methods & NMSE (dB) \\
            \hline
            \hline
            \multirow{3}{*}{1/16}
            & CRNet (Teacher) & Vanilla & $-$10.39 \\
            & CRNet-SE (Baseline) & Vanilla & $-$8.50 \\
            & CRNet-SE (Student) & Autoencoder KD & \textbf{$-$9.54} \\
            \hline
            \multirow{3}{*}{1/32}
            & CRNet (Teacher) & Vanilla & $-$7.90 \\
            & CRNet-SE (Baseline) & Vanilla & $-$7.11 \\
            & CRNet-SE (Student) & Autoencoder KD & \textbf{$-$7.31} \\
            \hline
            \hline

    \end{tabular}}
    \vspace{-0.1cm}
    \label{table:kd}	
\end{table}

\subsubsection{Inference Computational Complexity Analysis}
The adopted autoencoder networks mainly consist of convolutional layers, dense layers, batch normalization layers, and activation layers. In the following discussion, we ignore the complexity of the batch normalization and activation layers since it is negligible compared to that of the convolutional and dense layers. The floating point of operations (FLOPs) of convolutional layers \cite{molchanov2016pruning} are calculated as follows:
\begin{equation}
    {\rm FLOPs(Conv2D)} = 2H_{\rm out}W_{\rm out}(C_{\rm in}K_{\rm H}K_{\rm W}+1)C_{\rm out},
\label{equ:conv}
\end{equation}
where $C_{\rm in}$ and $C_{\rm out}$ denote the channel number of a convolutional layer input/output, respectively, and $H_{\rm out}$ and $W_{\rm out}$ are the height and width of a convolutional layer output, respectively. $K_{\rm H}$ and $K_{\rm W}$ denote the height and width of a convolutional kernel. The FLOPs of dense layer are calculated as follows:
\begin{equation}
    {\rm FLOPs(Dense)} = (2L_{\rm in}-1)L_{\rm out},
\label{equ:dense}
\end{equation}
where $L_{\rm in}$ and $L_{\rm out}$ represent the length of a dense layer input/output, respectively. Based on (\ref{equ:conv}) and (\ref{equ:dense}), the FLOPs of CRNet and CRNet-SE are listed in Table \ref{table:complexity}. When the compression ratio is 1/16 and 1/32, the FLOPs of CRNet-SE are 5.28\% and 3.05\% of CRNet FLOPs, respectively, demonstrating a significant reduction in encoder computational complexity.

\begin{table}[t]
	\vspace{0.7cm}
	\setlength{\abovecaptionskip}{-0.02cm}
	\setlength{\belowcaptionskip}{-0.4cm}	
	\centering
	\caption{Inference computational complexity of student and teacher autoencoder networks}	
	{\begin{tabular}[l]{c|c c c c c}					
			
			\hline
			\hline
			CR & Networks & UE FLOPs & Encoder FLOPs ratio\\
			\hline
			\multirow{2}{*}{1/16} & CRNet& 11,407,232 & \multirow{2}{*}{5.28\%}\\
                & CRNet-SE & 601,984 &
			 \\
			\hline
			\multirow{2}{*}{1/32} & CRNet & 11,145,152 & \multirow{2}{*}{3.05\%}\\
			& CRNet-SE & 339,904 &\\
			\hline
			\hline
	\end{tabular}}
	\vspace{-0.1cm}
	\label{table:complexity}	
\end{table}

Although FLOPs are useful reference when considering the computational complexity, network architecture and hardware also affect the real inference time. To achieve more accurate inference time, we conduct simulations on a Raspberry Pi 4 with a Cortex-A72 CPU and 8 GB memory, which has similar computational resources to most UE. The results are summarized in Table \ref{table:respberry}. The inference time of the student encoder is only 13.80\% to 14.76\% of the teacher encoder and less than 1 ms. Therefore, the lightweight encoder in CRNet-SE is suitable to be deployed in computation-limited devices in actual deployment.

\begin{table}[t]
    \vspace{0.7cm}
    \setlength{\abovecaptionskip}{-0.02cm}
    \setlength{\belowcaptionskip}{-0.4cm}	
    \centering
    \caption{Inferrence time of the encoder networks on Raspberry Pi}	
    {\begin{tabular}[l]{c | c c c}					
            \hline
            \hline
            CR & Networks & Inference time of encoder (ms) & Inference time ratio \\
            \hline
            \hline
            \multirow{2}{*}{1/16}
            & CRNet &  2.681 & \multirow{2}{*}{14.76\%}\\
            & CRNet-SE & 0.396 & \\
            \hline
            \multirow{2}{*}{1/32}
            & CRNet &   2.638 & \multirow{2}{*}{13.80\%}\\
            & CRNet-SE &  0.364 &\\
            \hline
            \hline

    \end{tabular}}
    \vspace{-0.1cm}
    \label{table:respberry}	
\end{table}

\subsubsection{Influence of Hyperparameters}
Simulations are conducted to investigate the influence of the hyperparameters $\alpha$ and $T$ in (\ref{equ:kd}) with a compression ratio of 1/16. First, autoencoder KD is performed with different $\alpha$ values, which control the balance between distillation loss and MSE loss. As shown in Table \ref{table:alpha}, the performance of autoencoder KD reaches the optimal when $\alpha$ is approximately 0.3. When $\alpha$ becomes larger, the weight of MSE loss increases, and autoencoder KD becomes more similar to directly training the student autoencoder. When $\alpha$ becomes smaller, the weight of distillation loss increases. However, using only the cross-entropy distillation loss fails to guide the training of the student network, resulting in poor performance when $\alpha=0$.

\begin{table*}[t]
    \vspace{0.7cm}
    \setlength{\abovecaptionskip}{-0.02cm}
    \setlength{\belowcaptionskip}{-0.4cm}	
    \centering
    \caption{Performance with different hyperparameter $\alpha$}	
    {\begin{tabular}[l]{c | c c c c c c c c c c c}					
            \hline
            \hline
            $\alpha$ & 0.0 & 0.1 & 0.2 & 0.3 & 0.4 & 0.5 & 0.6 & 0.7 & 0.8 & 0.9 & 1.0 \\
            \hline
            NMSE (dB) & 28.62 & $-$9.24 & $-$9.46 & \textbf{$-$9.54} & $-$9.45 & $-$9.43 & $-$9.39 & $-$9.39 & $-$9.31 & $-$9.27 & $-$8.50 \\
            \hline
            \hline
    \end{tabular}}
    \vspace{-0.1cm}
    \label{table:alpha}	
\end{table*}

The performance is also tested with different $t$ values, which determine the smoothness of the extended softmax function in (\ref{equ:softmaxt}). As shown in Table \ref{table:T}, the student autoencoder achieves the best performance when $t$ is around 5. When $t=1$, the extended softmax function degrades to the ordinary softmax function, which fails to enhance the dark knowledge effectively. When $t$ becomes larger, the output of the extended softmax output is closer to uniform. An appropriate $t$ can effectively enhance the dark knowledge, but too large $t$ can lead to information loss because all output values are almost the same.

\begin{table*}[t]
    \vspace{0.7cm}
    \setlength{\abovecaptionskip}{-0.02cm}
    \setlength{\belowcaptionskip}{-0.4cm}	
    \centering
    \caption{Performance with different hyperparameter $t$}
    {\begin{tabular}[l]{c | c c c c c c c c c c}					
            \hline
            \hline
            $t$ & 1 & 2 & 3 & 4 & 5 & 6 & 7 & 8 & 9 & 10 \\
            \hline
            NMSE (dB) & $-$8.31 & $-$9.11 & $-$9.36 & $-$9.43 & \textbf{$-$9.54} & $-$9.40 & $-$9.27 & $-$9.25 & $-$9.13 & $-$9.00 \\
            \hline
            \hline
    \end{tabular}}
    \vspace{-0.1cm}
    \label{table:T}	
\end{table*}

\subsection{Encoder KD for CSI Feedback}
\subsubsection{Performance Analysis}
The performance of encoder KD is then evaluated for different compression ratios, as shown in Table \ref{table:ekd}. The results demonstrate that encoder KD can further improve the feedback performance compared to the autoencoder KD-based method. In addition, the performance of the intellectual property-protecting encoder KD method (decoder fine-tuning) is tested. Fine-tuning only the decoder results in negligible performance drop compared to fine-tuning the whole autoencoder and still outperforms the autoencoder KD-based method. These results indicate that the intellectual property-protecting encoder KD can successfully improve the performance of the student autoencoder.

\begin{table}[t]
    \vspace{0.7cm}
    \setlength{\abovecaptionskip}{-0.02cm}
    \setlength{\belowcaptionskip}{-0.4cm}	
    \centering
    \caption{NMSE comparison among encoder KD with other training strategies}	
    {\begin{tabular}[l]{c| c c c}					
            \hline
            \hline
            CR & Networks & Methods & NMSE (dB)\\
            \hline

            \hline
            \multirow{5}{*}{1/16}
            & CRNet (Teacher) & Vanilla & $-$10.39 \\
            & CRNet-SE (Baseline) & Vanilla & $-$8.50 \\
            & CRNet-SE (Student) & Autoencoder KD & $-$9.54 \\
            & CRNet-SE (Student) & Encoder KD & \textbf{$-$9.75} \\
            & CRNet-SE (Student) & Variant encoder KD & $-$9.61 \\
            \hline
            \multirow{5}{*}{1/32}
            & CRNet (Teacher) & Vanilla & $-$7.90 \\
            & CRNet-SE (Baseline) & Vanilla & $-$7.11 \\
            & CRNet-SE (Student) & Autoencoder KD & $-$7.31 \\
            & CRNet-SE (Student) & Encoder KD & \textbf{$-$7.48} \\
            & CRNet-SE (Student) & Variant encoder KD & $-$7.45 \\
            \hline
            \hline

    \end{tabular}}
    \vspace{-0.1cm}
    \label{table:ekd}	
\end{table}

\subsubsection{Training Computational Complexity Analysis}
The computational complexity of the training process in autoencoder KD and encoder KD-based methods is compared. First, both methods include pretraining the teacher autoencoder, which can be completed by offline training and not considered in training complexity. Then, two different KD methods are performed. For autoencoder KD, the whole student autoencoder is trained from scratch. For encoder KD, only the student encoder is trained from scratch and the pretrained decoder network is directly used. Therefore, the computational complexity of the distillation process in encoder KD is significantly lower than in autoencoder KD because the training complexity of the student encoder for one epoch is much smaller compared with that of the whole autoencoder. In addition, the student encoder can be trained with fewer epochs due to its simple architecture. Third, the encoder KD method requires end-to-end fine-tuning to further improve the performance. This step only takes a small number of epochs and does not introduce much training complexity. Let $n_{\rm au,1}$, $n_{\rm en,1}$, and $n_{\rm en,2}$ be the epoch numbers of autoencoder KD, encoder KD, and end-to-end fine-tuning, respectively, $c_{\rm en,s}$, $c_{\rm de,s}$, and $c_{\rm de,t}$ be the FLOPs of student encoder, student decoder, teacher decoder. The training time of autoencoder KD is proportional to  
\begin{equation}
   t_{\rm au} \propto n_{\rm au,1}(c_{\rm en,s}+c_{\rm de,s}),
\end{equation}
and the training time of encoder KD is proportional to
\begin{equation}
    t_{\rm en} \propto n_{\rm en,1}c_{\rm en,s}+n_{\rm en,2}(c_{\rm en,s}+c_{\rm de,t}).
\end{equation}
As analyzed above, although encoder KD requires an additional fine-tuning step, the training time of encoder KD is still much smaller than autoencoder KD, that is, $t_{\rm au}\gg t_{\rm en}$. The reduction in training time is mainly attributed to the direct use of pretrained teacher decoder.

Table \ref{table:trainingtime} shows the training time of encoder KD and other training methods with a compression ratio of 1/16.\footnote{The simulation results indicate that the training time for different compression ratios is close to each other.} ``Distillation and direct training'' refers to training the entire autoencoder network from scratch (for CRNet and CRNet-SE) or the distillation process (for CRNet-SE KD and CRNet-SE Encoder KD). ``Fine-tuning'' denotes the end-to-end fine-tuning for encoder KD. As shown in Table \ref{table:trainingtime}, training CRNet and CRNet-SE from scratch takes 3.88 and 3.64 hours, respectively. Performing autoencoder KD from CRNet to CRNet-SE takes 3.71 hours, which is similar to training an autoencoder from scratch. The long training time is mainly due to the complicated architectures of decoder networks. For encoder KD, only a lightweight encoder network is trained from scratch, which takes 1.23 hours and is significantly faster than autoencoder KD. Although end-to-end fine-tuning is performed in encoder KD, the fine-tuning process can be completed in a small number of epochs and does not take much training time compared with the distillation stage. The total training time can be reduced by 66.8$\%$ compared with autoencoder KD. In addition, simulations show that when the fine-tuning time is reduced from 0.32 h to 0.05 h in encoder KD, the performance of encoder KD can still outperform directly training the student network CRNet-SE. This means the fine-tuning time can be further reduced if small performance loss is acceptable.

\begin{table*}[t]
    \vspace{0.7cm}
    \setlength{\abovecaptionskip}{-0.02cm}
    \setlength{\belowcaptionskip}{-0.4cm}	
    \centering
    \caption{Training time comparison between encoder KD with other training strategies}	
    {\begin{tabular}[l]{c c c c c}					
            \hline
            \hline
            Networks & Methods & Distillation or direct training & Fine-tuning & Total \\
            \hline
            CRNet (Teacher) & Vanilla & 3.88 h & 0 h & 3.88 h\\
            CRNet-SE (Baseline) & Vanilla & 3.64 h & 0 h & 3.64 h\\
            CRNet-SE (Student) & Autoencoder KD & 3.71 h & 0 h & 3.71 h\\
            CRNet-SE (Student) & Encoder KD & \textbf{0.91 h} & 0.32 h & \textbf{1.23 h}\\
            \hline
            \hline
    \end{tabular}}
    \vspace{-0.1cm}
    \label{table:trainingtime}	
\end{table*}

\subsubsection{Generalization Improvement with Encoder KD}
As mentioned above, encoder KD can effectively transfer knowledge from a teacher network to a student network, which consequently boosts the feedback performance of the student network. However, previous simulations only focus on achieving high feedback performance in a target domain represented by a single dataset. In this part, we are interested in whether a well-generalized teacher network can be used to improve the generalization capability of a student network with encoder KD. This is motivated by the fact that the BS can usually access and store diverse CSI datasets due to sufficient storage resources. By mixing different pre-collected datasets into the training dataset of the teacher network, we can achieve a teacher network generalized to different scenarios (e.g., UMi, RMa, LOS, and NLOS), environments (different areas in a certain scenario), or settings (e.g., bandwidth). However, the BS may not be willing to directly share these pre-collected datasets with UEs due to privacy concerns or limited bandwidth. In this case, we attempt to use the intellectual property-protecting encoder KD to transfer well-generalized knowledge from the teacher network to the student network.\footnote{Due to intellectual property problems mentioned in Section IV, we consider the case where the BS and UE have no knowledge of the network architectures of each other in this part. Therefore, only intellectual property-protecting encoder KD is considered in this part.}

The performance of the encoder KD method in improving generalization is then tested. Initially, the encoder KD method is used to enhance generalization across different scenarios by increasing the width of convolutional layers in CRBlock from 8 to 16 for better performance. Datasets 1 and 2 are generated in UMi LOS and UMi NLOS, respectively. We assume a UE in the UMi LOS scenario that collects Dataset 1 and shares it with the BS. The BS expects a well-generalized student network to the UMi NLOS scenario, and hence trains a teacher network with a combination of Dataset 1 and 2 for variant encoder KD (Step 1).
The teacher network is then used to apply variant encoder KD to the student network CRNet-SE with only Dataset 1 (Step 2 and 3). We also train a CRNet-SE by the vanilla training method with only Dataset 1 as a baseline.

Table \ref{table:scenario} shows that even though the UE has no access to Dataset 2, the BS and UE still cooperate to train a student network that is well-generalized to the UMi NLOS scenario. Variant encoder KD brings a 2.01 dB performance gain compared with the baseline. Notably, the performance of the student network after variant encoder KD and the teacher network is slightly lower than the baseline network when evaluated on Dataset 1. This is because the former two networks have to learn CSI compression and reconstruction in more complicated scenarios, i.e., both UMi LOS and UMi NLOS scenarios, while the baseline network only needs to fit the UMi LOS scenario, resulting in slightly overfitting on Dataset 1. However, significantly improving the performance on Dataset 2 with minor performance drop on Dataset 1 is still reasonable.

\begin{table*}[t]
    \vspace{0.7cm}
    \setlength{\abovecaptionskip}{-0.02cm}
    \setlength{\belowcaptionskip}{-0.4cm}	
    \centering
    \caption{Generalization between different scenarios}
    {\begin{tabular}[l]{c  c  c |c c}					
            \hline
            \hline
            Networks & Methods & \diagbox{Train}{Test} & Dataset 1 & Dataset 2\\
            \hline
            CRNet (Teacher) & Vanilla & Dataset 1\&2 & $-$10.18 & $-$6.35\\
            CRNet-SE (Baseline) & Vanilla & Dataset 1 & $-$10.3 & $-$2.83 \\
            CRNet-SE (Student) & Variant encoder KD & Dataset 1 & $-$9.91 & $-$4.84\\
            \hline
            \multicolumn{3}{c|}{Gain} & $-$0.39 & 2.01 \\
            \hline
            \hline
    \end{tabular}}
    \vspace{-0.1cm}
    \label{table:scenario}	
\end{table*}

Then, similar simulations are conducted on different environments. Dataset 1 and Dataset 2 are generated in different areas in the UMi NLOS scenario, and the radius of each area is 10 m. The results are summarized in Table \ref{table:environment}. The variant encoder KD method also brings a 2.48 dB performance gain on Dataset 2. This means the generalization between different environments can be effectively improved with the proposed variant encoder KD method.

\begin{table*}[t]
    \vspace{0.7cm}
    \setlength{\abovecaptionskip}{-0.02cm}
    \setlength{\belowcaptionskip}{-0.4cm}	
    \centering
    \caption{Generalization between different environments}	
    {\begin{tabular}[l]{c  c  c |c c}					
            \hline
            \hline
            Networks & Methods & \diagbox{Train}{Test} & Dataset 1 & Dataset 2\\
            \hline
            CRNet (Teacher) & Vanilla & Dataset 1\&2 & $-$6.95 & $-$8.7\\
            CRNet-SE (Baseline) & Vanilla & Dataset 1 & $-$6.53 & $-$3.73 \\
            CRNet-SE (Student) & Variant encoder KD & Dataset 1 & $-$6.30 & $-$6.21\\
            \hline
            \multicolumn{3}{c|}{Gain} & $-$0.23 & 2.48 \\
            \hline
            \hline
    \end{tabular}}
    \vspace{-0.1cm}
    \label{table:environment}	
\end{table*}

The simulations are conducted on the dataset generated with different bandwidths in an Indoor LOS scenario. Dataset 1 and 2 adopt 40 MHz and 10 MHz bandwidth, respectively, while the numbers of subcarriers are both set to 32. As shown in Table \ref{table:bandwidth}, the performance on Dataset 2 improves from $-$8.51 to $-$11.15 dB, which proves that the variant encoder KD method can improve the generalization capability between different bandwidths.

\begin{table*}[t]
    \vspace{0.7cm}
    \setlength{\abovecaptionskip}{-0.02cm}
    \setlength{\belowcaptionskip}{-0.4cm}	
    \centering
    \caption{Generalization between different bandwidths}	
    {\begin{tabular}[l]{c  c  c |c c}					
            \hline
            \hline
            Networks & Methods & \diagbox{Train}{Test} & Dataset 1 & Dataset 2\\
            \hline
            CRNet (Teacher) & Vanilla & Dataset 1\&2 & $-$12.76 & $-$16.08\\
            CRNet-SE (Baseline) & Vanilla & Dataset 1 & $-$11.15 & $-$8.51 \\
            CRNet-SE (Student) & Variant encoder KD & Dataset 1 & $-$11.10 & $-$11.15\\
            \hline
            \multicolumn{3}{c|}{Gain} & -0.05 & 2.64 \\
            \hline
            \hline
    \end{tabular}}
    \vspace{-0.1cm}
    \label{table:bandwidth}	
\end{table*}

\subsubsection{Comparison with Sequential Training}

As discussed in Section IV, a sequential training method is introduced to safeguard intellectual property for various vendors \cite{qualcomm}.\footnote{To ensure a fair comparison, we adopt the BS-first sequential training approach \cite{38843}. Additionally, we consider a scenario where the encoder networks at the BS and UE share the same lightweight architecture, specifically, the encoder of CRNet-SE. This configuration is commonly adopted in the evaluation of separate training methods by most companies within 3GPP \cite{38843}. } In our comparison, we evaluate the performance of sequential training against the proposed intellectual property-protecting encoder KD-based method. According to Table \ref{table:st}, the intellectual property-protecting encoder KD method demonstrates superior performance compared to sequential training. Sequential training with CRNet-SE fails to achieve high feedback performance due to the direct utilization of a lightweight encoder. On the other hand, the proposed encoder KD method can be employed to further enhance the feedback performance in real-world deployment.

\begin{table}[t]
    \vspace{0.7cm}
    \setlength{\abovecaptionskip}{-0.02cm}
    \setlength{\belowcaptionskip}{-0.4cm}	
    \centering
    \caption{NMSE comparison among sequential training with other training strategies}	
    {\begin{tabular}[l]{c| c c c}					
            \hline
            \hline
            CR & Networks & Methods & NMSE (dB)\\
            \hline

            \hline
            \multirow{6}{*}{1/16}
            & CRNet (Teacher) & Vanilla & $-$10.39 \\
            & CRNet & Sequential training \cite{qualcomm} & $-$10.25 \\
            & CRNet-SE  & Vanilla & $-$8.50 \\
            & CRNet-SE  & Sequential training \cite{qualcomm} & $-$8.39 \\
            & CRNet-SE (Student) & Variant encoder KD & $-$9.61 \\
            \hline
            \multirow{6}{*}{1/32}
            & CRNet (Teacher) & Vanilla & $-$7.90 \\
            & CRNet  & Sequential training \cite{qualcomm} & $-$7.81 \\
            & CRNet-SE  & Vanilla & $-$7.11 \\
            & CRNet-SE  & Sequential training \cite{qualcomm} & $-$6.97 \\
            & CRNet-SE (Student) & Variant encoder KD & $-$7.45 \\
            \hline
            \hline

    \end{tabular}}
    \vspace{-0.1cm}
    \label{table:st}	
\end{table}

\section{Conclusion}
This paper introduces KD as a method to realize lightweight autoencoder networks in DL-based CSI feedback, which is compatible with other lightweight methods. Initially, an autoencoder KD method is proposed, where knowledge learned by a complex teacher autoencoder is transferred to a lightweight student autoencoder, thereby improving the performance of the student autoencoder. To further reduce the training time required for KD, an encoder KD method is introduced. In this approach, the KD process is exclusively applied to the student encoder. Additionally, we adapt the encoder KD method to create a variant that protects intellectual property. This variant prevents the sharing of decoder (or encoder) architectures between the BS (or UE) and UE (or BS), respectively. The performance of these KD-based neural network lightweight methods is evaluated using channel datasets generated by the QuaDRiGa channel model. The simulation results demonstrate that the feedback performance of lightweight autoencoder networks can be significantly improved through autoencoder KD, leveraging knowledge from more complex teacher networks. Encoder KD further enhances feedback performance while reducing training time, and the variant encoder KD method safeguards the intellectual property of different vendors with minimal impact on performance. Moreover, the variant encoder KD-based method effectively enhances the generalization capability of student networks across various scenarios, environments, and settings. Given the encouraging outcomes of this research, we anticipate that our work will stimulate further exploration into the application of KD-based lightweight technology in other physical layer domains.

\bibliographystyle{IEEEtran}
\bibliography{refer1}

\end{document}